\documentclass[letters,a4paper,fleqn,usenatbib]{mnras}

\usepackage{newtxtext,newtxmath}

\usepackage[T1]{fontenc}
\usepackage{ae,aecompl}

\usepackage{graphicx}	
\usepackage{amsmath}	
\usepackage{amssymb}	
\usepackage{hyperref}
\usepackage{pdflscape}
\usepackage{wasysym}

\newcommand{\degree}{{^{\circ}}}

\title[The 2019 Taurid resonant swarm]{The 2019 Taurid resonant swarm: prospects for ground detection of small NEOs }

\author[D. Clark et al.]{
David Clark,$^{1,2,3}$\thanks{E-mail: dclark56@uwo.ca}
Paul Wiegert,$^{2,3}$
Peter G. Brown$^{2,3}$
\\
$^{1}$Department of Earth Sciences, University of Western Ontario, London, Ontario, N6A 5B7, Canada\\
$^{2}$Department of Physics and Astronomy, University of Western Ontario, London, Ontario, N6A 3K7, Canada\\
$^{3}$Centre for Planetary Science and Exploration, University of Western Ontario, N6A 5B7, Canada\\
}

\date{Accepted XYZ; Received XYZ; in original form XYZ}

\pubyear{2019}

\begin{document}
\label{firstpage}
\pagerange{\pageref{firstpage}--\pageref{lastpage}}
\maketitle

\begin{abstract}
In June 2019 the Earth will approach within 5$^\circ$  mean anomaly of the centre of the Taurid resonant swarm, its closest post-perihelion encounter with Earth since 1975. This will be the best viewing geometry to detect and place limits on the number of NEOs proposed to reside at the swarm centre until the early 2030s. We present an analysis of the optimal times and pointing locations to image NEOs associated with the swarm.
\end{abstract}

\begin{keywords}
minor planets, asteroids, general -- meteorites, meteors, meteoroids -- comets: individual:2P/Encke
\end{keywords}



\section{Introduction}

The Taurid meteor shower has long been recognised as highly unusual \citep{Whipple1967}. Its peculiar properties include its long duration ($\sim$6 months) \citep{Stohl1986}, dispersed radiant and presence of large (meter-sized) particles \citep{Spurny2017}. The Daytime beta-Taurids peak in late June and the Tunguska event has been suggested as originating from this shower \citep{Kresak1978}. Additionally, the main showers (the North and South Taurids) which occur in late autumn are dynamically related to a broader complex of many showers occurring throughout the year \citep{Stohl1990} . These are all linked to the unusual comet 2P/Encke. The large mass of material associated with the stream and its apparent dominance of the sporadic complex as a whole \citep{Stohl1986} led to development of the Taurid complex -- giant comet hypothesis by \cite{Clube1984}. 

This hypothesis proposes that a giant comet (of order one hundred kilometres -- comparable to large KBOs) fragmented in the inner solar system of order 10-20 ka ago, producing a complex of dust and small bodies (including 2P/Encke and associated asteroids) still present today. 
\cite{Asher1993b} further developed this Taurid complex hypothesis, proposing that many near-Earth Objects related to this breakup are trapped in the 7:2 MMR with Jupiter. They termed this concentration of material the Taurid Resonant Swarm (TS).  As a consequence, \cite{Asher1993a} and\cite{Asher1993b} proposed that at certain epochs when Earth passes close to the centre of this resonance  (and hence the hypothesised TS) we should detect an enhancement in both dust and larger objects associated with the Taurids in near-Earth space, including impacts on Earth \citep{Asher1994}. Among the evidence supporting the existence of the Taurid Swarm is increased fireball activity near the time of TS swarm passages close to Earth \citep{Asher1998, Beech2004a} and a large increase in seismically detected lunar impacts near the time of the 1975 TS encounter \citep{Oberst1987, Asher1993a}. 

This hypothesis has been more broadly adapted to a school of thought which posits that larger NEO impacts (tens of meters to hundreds of meter sized objects) at Earth are correlated with the Taurids. A consequence of this is that we might expect a large fraction of impacts to be coherent (rather than random) in time. This ``Coherent Catastophism'' school \citep{Asher1994} is not universally accepted, some suggesting that the orbits occupied by the Taurids lead naturally to long dynamical lifetimes and potentially spurious orbital similarities among NEAs \citep{Valsecchi1995} and supported by the observation that many NEAs associated previously with the Taurid complex \citep{Asher1993c, Steel1996} do not show similar reflectance spectra \citep{Snodgrass2015}, particularly among the largest members \citep{Cristescu2014}. 

While evidence for the TS has been slowly accumulating for some years, the most recent TS encounter in 2015 has provided some of the most convincing data to date. 

During the 2015 encounter, a large increase in fireball activity associated with the Taurids as predicted by the TS theory \citep{Asher1998} was very well documented \citep{Spurny2017}. Over ~100 bright Taurids were reported by \citep{Spurny2017} using high precision fireball cameras. They showed for the first time that much of the increased Taurid activity was confined to a radiant branch associated with the 7:2 MMR as predicted by \cite{Asher1993a}.

Interestingly, several asteroids having very similar orbits \citep{Olech2016, Olech2017} and a number of meter-sized fireballs were also recorded during the expected time of the TS encounter \citep{Spurny2017, Olech2016}. This is among the strongest quantitative dynamical evidence to date for the TS.

In 2019, Earth will encounter the TS again, with slightly better geometry than in 2015, this time on the outbound (post-perihelion) leg of the shower. Since the TS location as predicted by \cite{Asher1993a} is known, it is possible to directly image any TS NEOs in the tens of meter--hundreds of meter size range. This would provide a direct observational test of the TS hypothesis, most particularly whether a dense swarm of material as described recently by \cite{Napier2015} is present. 
\section{Taurid Swarm Simulation}   
In order to assess the optimum observation conditions, we construct a model swarm, which is done in a two-stage process. First, particles are generated with uniformly random orbital elements across the ranges discussed below, then the particles are integrated forward in time from near the present epoch for 1000 years. The resonant argument for the 7:2 mean motion resonance
\begin{equation}
\phi = (p+q)\lambda_p- p\lambda - q\varpi 
\end{equation}
where $p=2$, $q=5$, $\lambda_p$ is the mean longitude of Jupiter, $\lambda$ is the mean longitude of the particle, and $\varpi$ is the particle's longitude of perihelion, is computed over this interval, and non-resonant particles are discarded. \cite{Asher1998} report that enhancements in the Taurid fireball flux is associated with the Earth passing within $\pm 40\deg$ of the centre of resonance. However, a cloud of this extent is large on the sky. So for the purpose of devising an efficient observational strategy, we concentrate our attention on the 'core' of the hypothetical swarm by considering here only particles with $\phi$ within $6 \degree$ of the resonance centre (The full cloud is discussed later in section 4, but consideration of the full cloud adds little to the discussion of optimal targeted observing).

The ranges of semimajor axis used were $2.23$~AU$<a<2.27$~AU and for the longitude of perihelion $145\degree < \varpi < 165\degree$, and the resulting resonant amplitudes of the simulated particles are a strong function of these elements. But though the state of being in the 7:2 resonance constrains the semimajor axis $a$ and longitude of perihelion $\varpi$ of the swarm particles, it leaves the other orbital elements free to vary. We choose values for the orbital elements empirically based on those observed by (see fig. 12 in \cite{Spurny2017}) for fireballs associated with the 7:2 branch of the Taurids:  $0.25$~AU$< q < 0.45$~AU, $4.5\degree < i < 4.6\degree$, $35\degree < \Omega < 45\degree$. The orbits of the collection of particles generated from the orbital ranges and other properties discussed above constitute our model for the Taurid swarm for the assessment of observability during its near pass to Earth in 2019. We acknowledge that the choices of orbital elements favours the near Earth intersecting portion of a potentially much larger TS. This is done to enable a more focused observation strategy.
\begin{figure}
    \centering
    \fbox{\includegraphics[width=3.8cm]{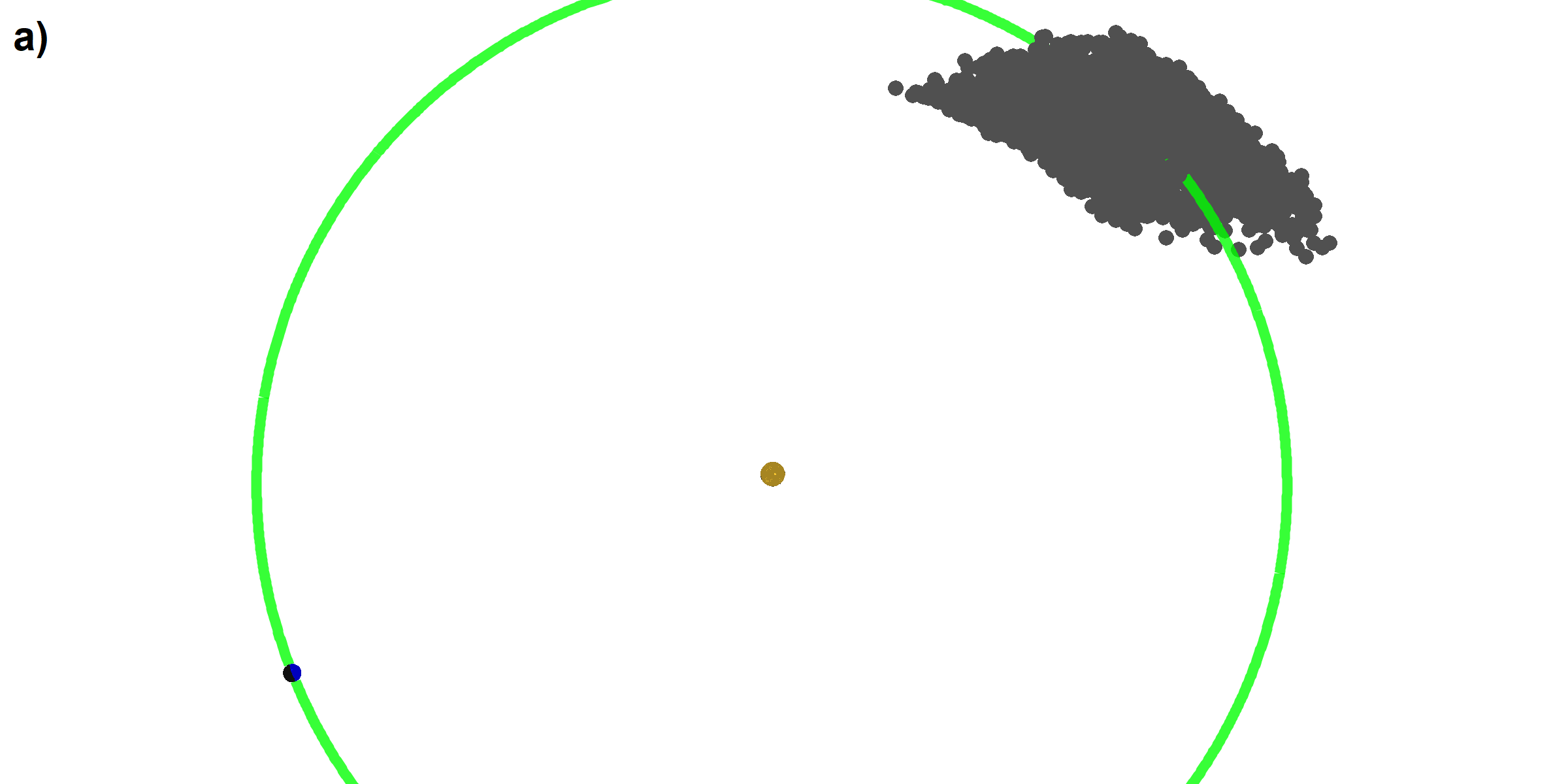}}
    \fbox{\includegraphics[width=3.8cm]{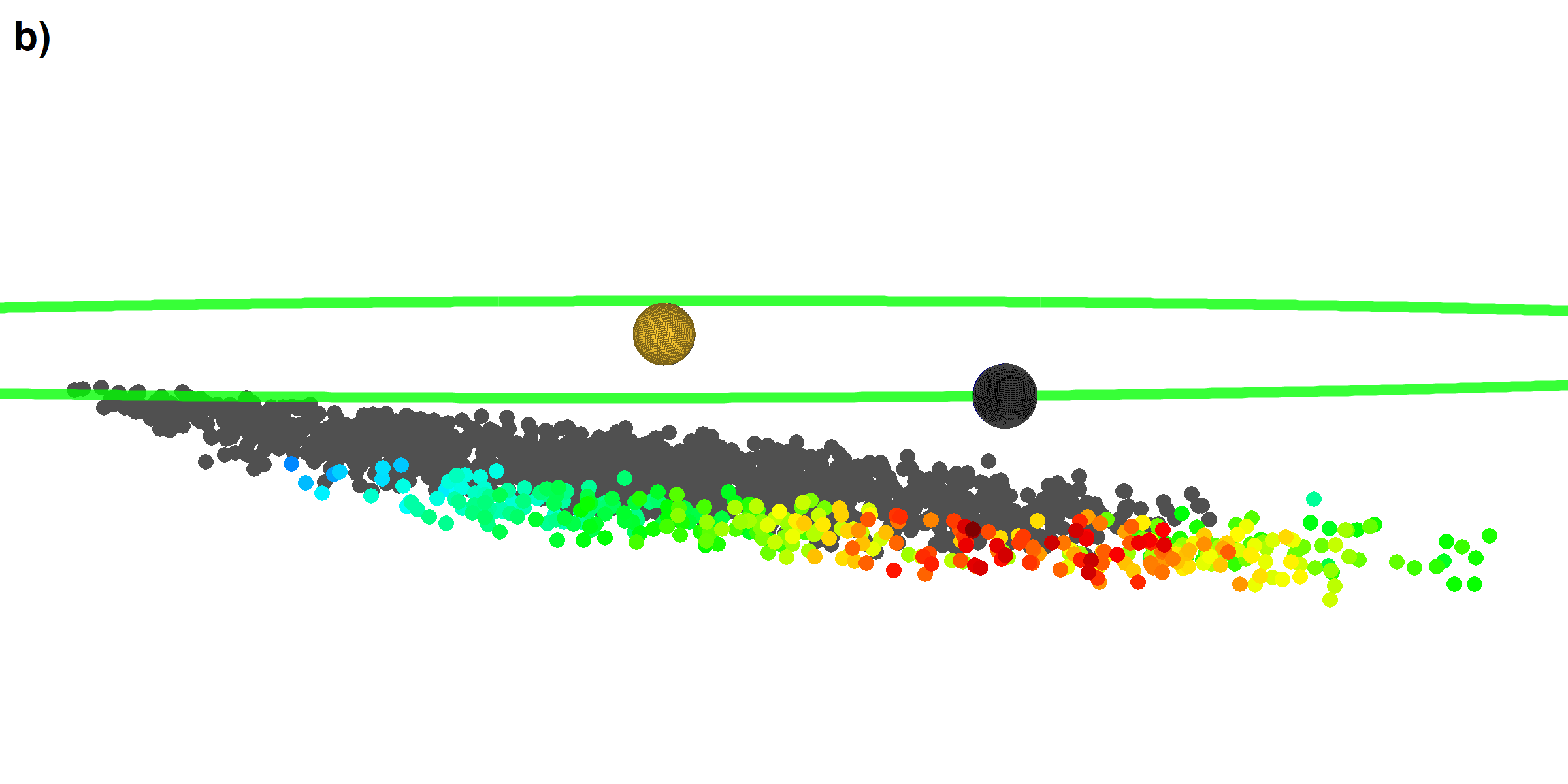}} 
    \fbox{\includegraphics[width=3.8cm]{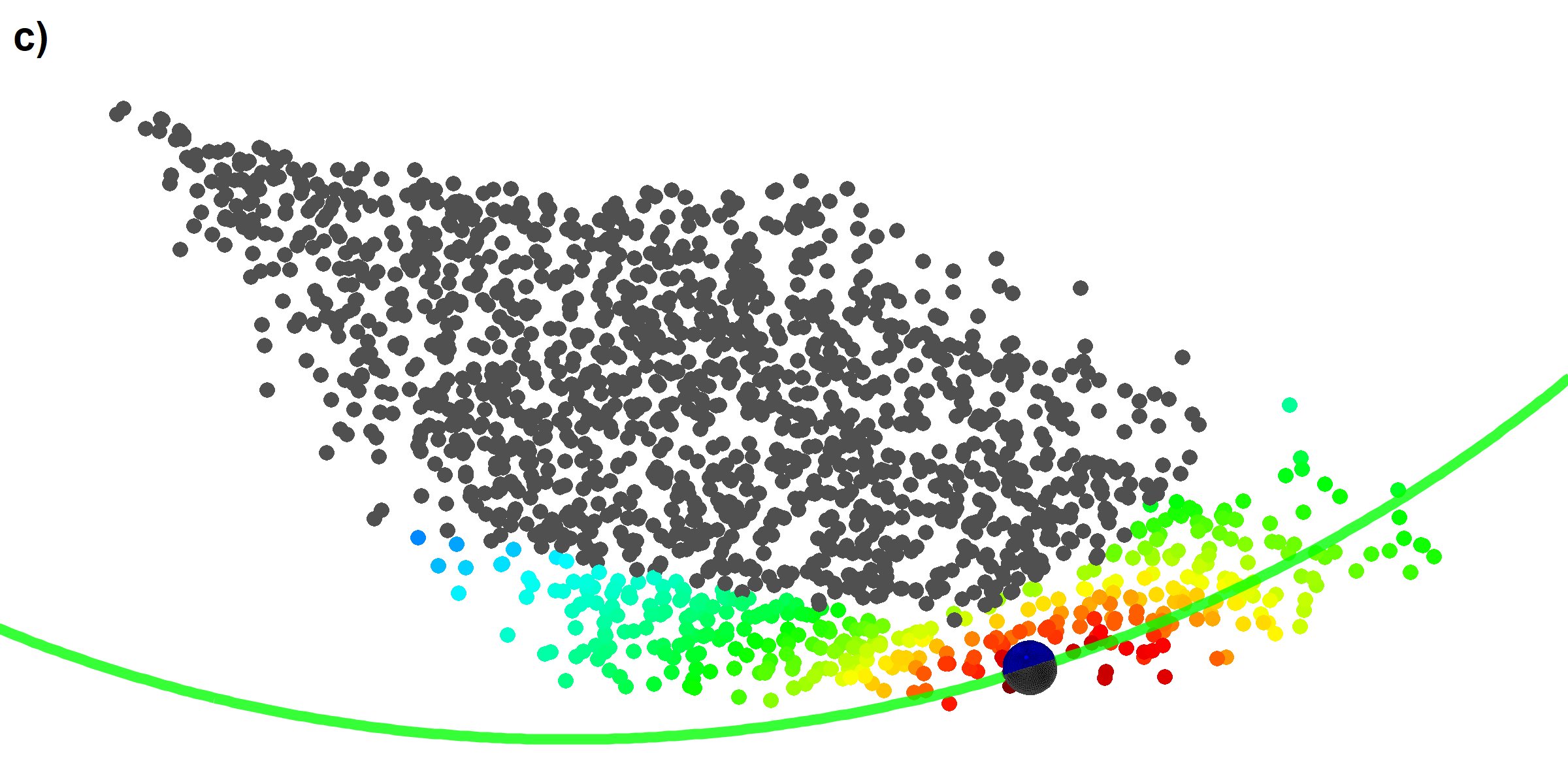}} 
    \fbox{\includegraphics[width=3.8cm]{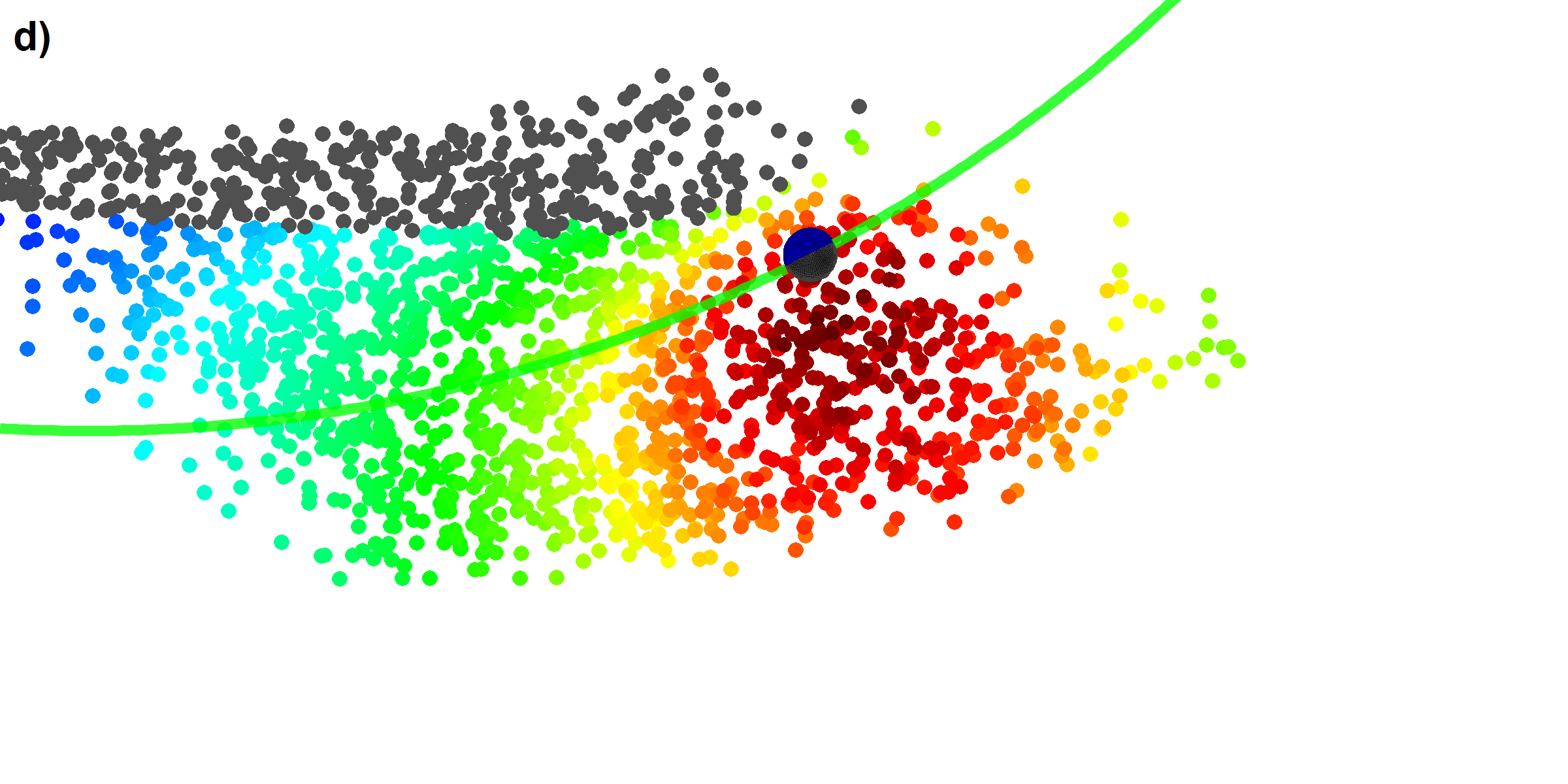}} 
    \fbox{\includegraphics[width=3.8cm]{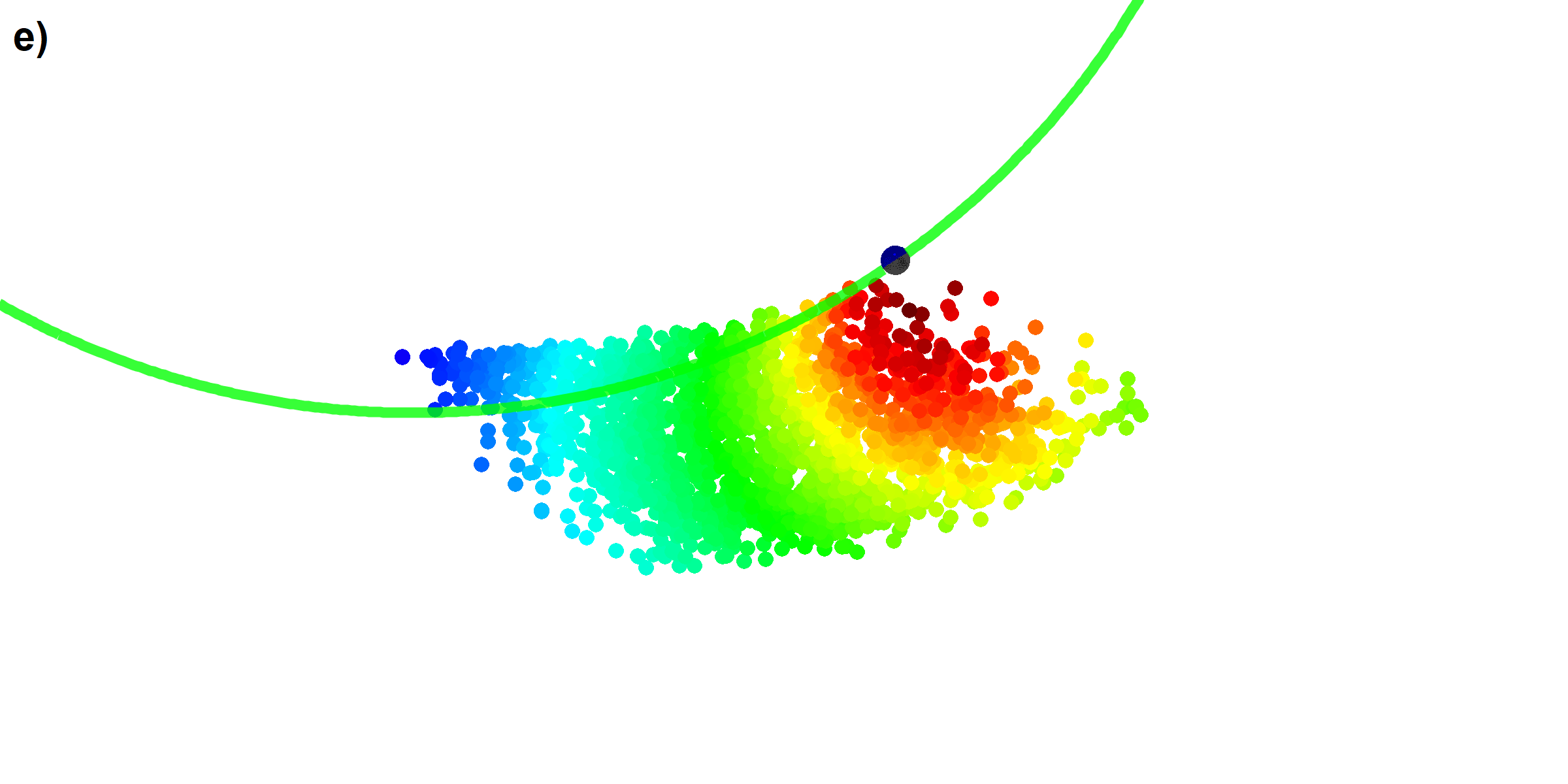}} 
    \fbox{\includegraphics[width=3.8cm]{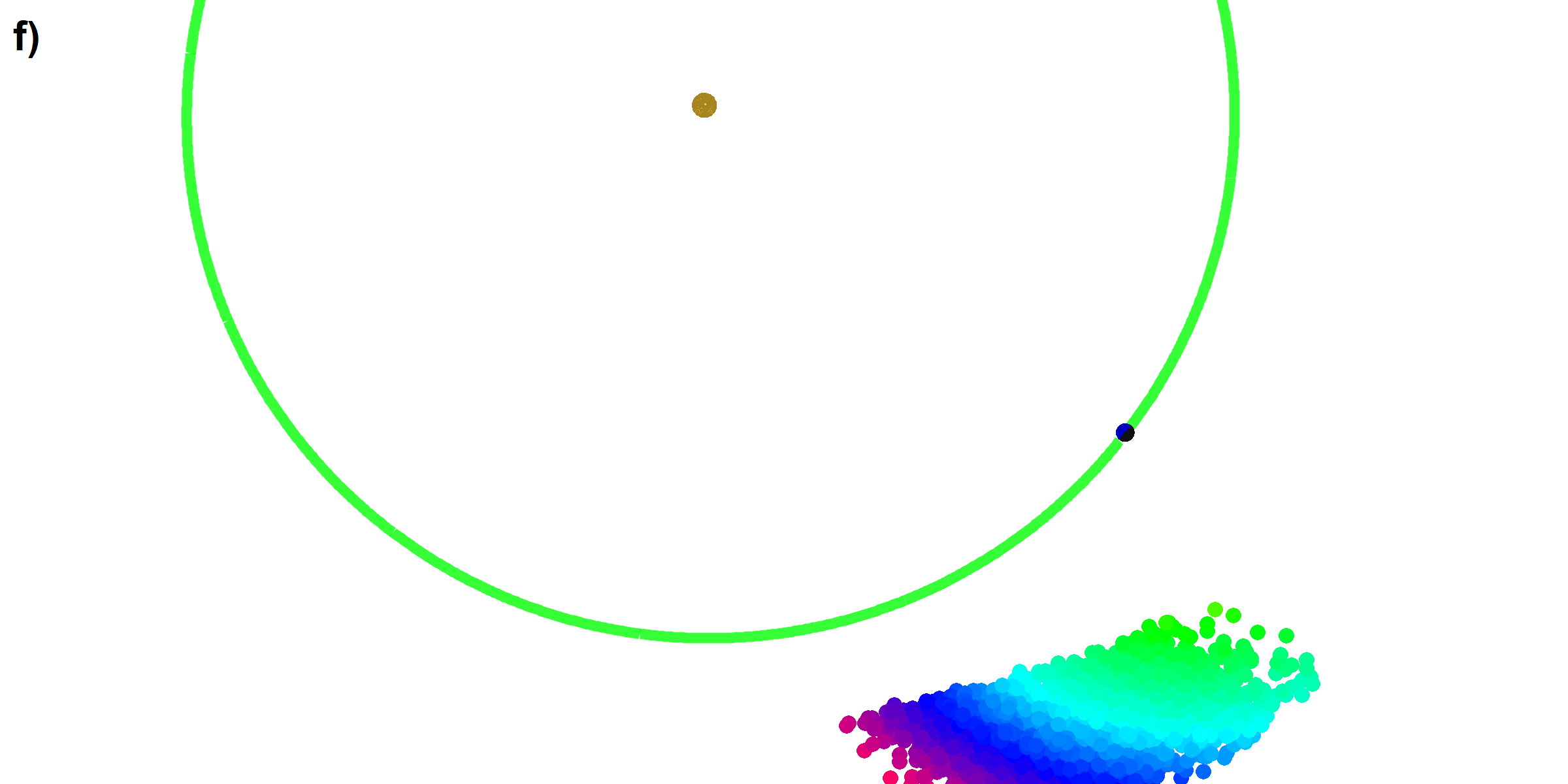}} 
    \fbox{\includegraphics[width=3.8cm]{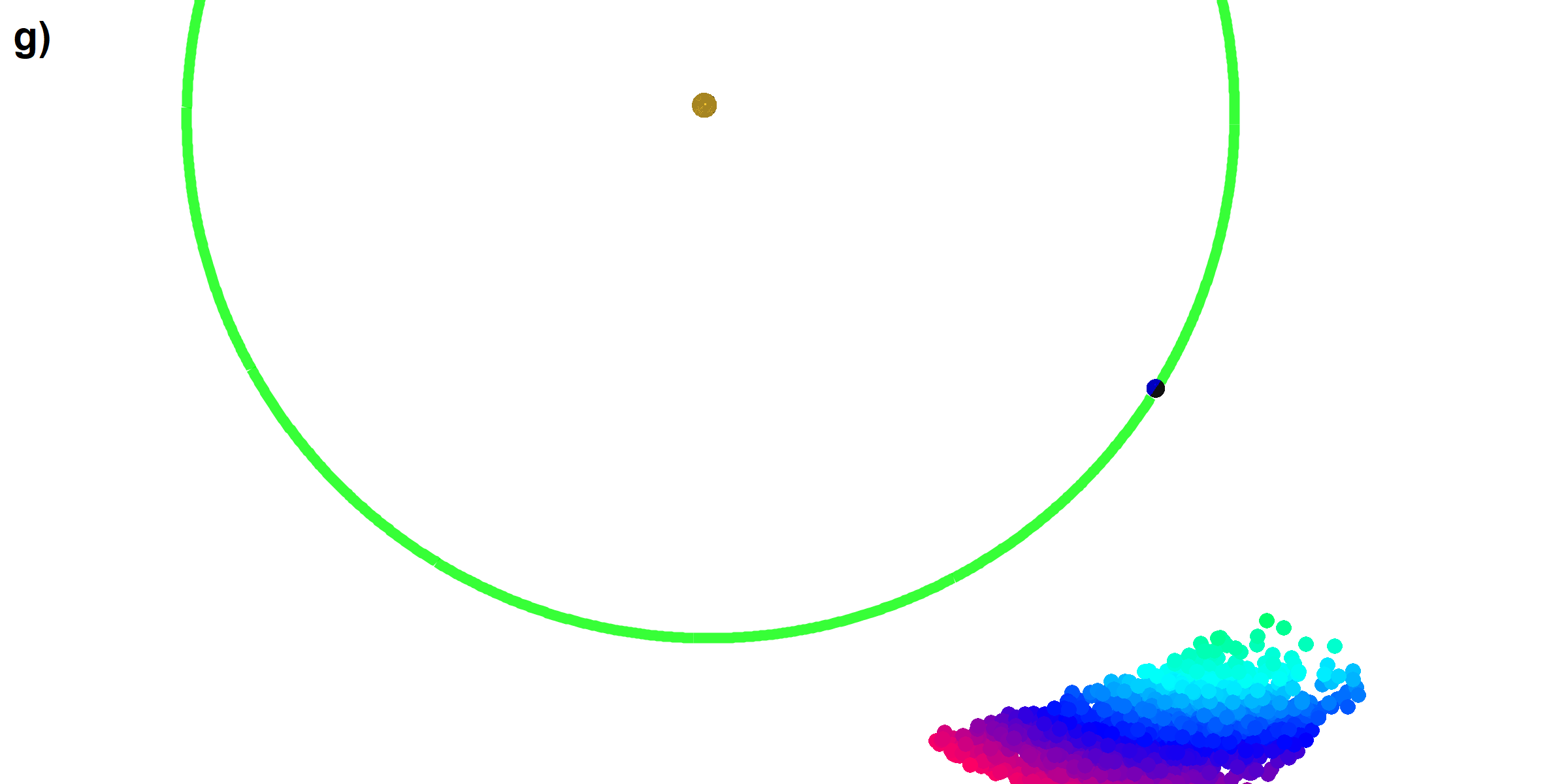}} 
    \fbox{\includegraphics[width=3.8cm]{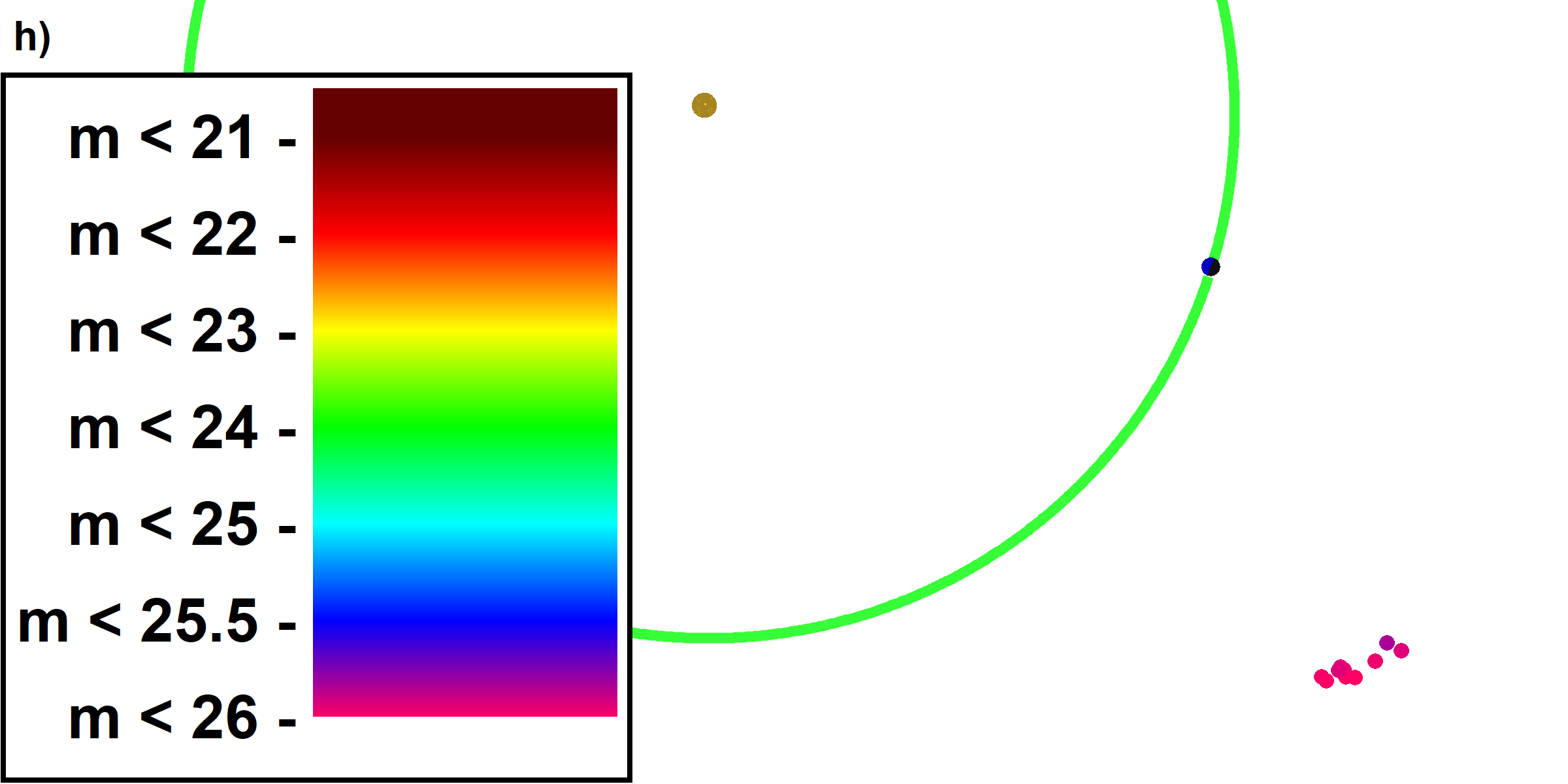}} 
    \caption{TSC viewing circumstances progressing in time from box a) through h). a)~Swarm inbound during April 2019. b)~Swarm passing Earth below the ecliptic. c),d),e)~swarm passing beneath the Earth June 26, July 04, and July 14 respectively.  f),g)~Decreased apparent magnitudes by August 1 and 7 respectively. h)~All $H<24$ objects lost from visibility by August 24. d) and g) correspond to the midpoints of viewing opportunities A and B referred to in Fig.\ref{Fig:MagAndMotion} and the text. All TSC objects have an assumed $H~~24$.  Object colours represent apparent magnitudes corresponding to the legend in box h). Objects with solar elongation $<~60\degree$ are coloured grey.  Objects of $m>26$ are not shown.}
    \label{Fig:Circumstances}
\end{figure}
\section{Observing Geometry and Visibility}
With the goal of this article being to assist in the discovery new large members of the TS, we are assuming a baseline absolute magnitude $H=24$ for all objects in our model TS.  This corresponds to a diameter of $100$ meters at a cometary albedo of $A=0.05$. This is consistent with known objects believed associated with the TS \citep{Spurny2017, Olech2016}.

The centre of the TS roughly follows the path of 2P/Encke, somewhat less inclined, with a width of approximately $30\degree$ in solar longitude at $1$~AU from the Sun, and a duration of Earth orbit passage of approximately one month. On the inbound leg, the TS core (or TSC) passes through the Earth's orbit throughout the month of April 2019 spanning solar longitudes of approximately $27\degree$ to $55\degree$, but at this time the Earth is on the opposite side of its orbit (Fig.~\ref{Fig:Circumstances}a).  The swarm passes rapidly though perihelion at $0.25-0.45$~AU, proceeding to an outbound near-Earth encounter from approximately 23 June - 17 July spanning solar longitudes $257-286\degree$. At this point in its orbit the TSC is approximately $0.06-0.10$~AU below the ecliptic (Fig.~\ref{Fig:Circumstances}b).  The Earth passes over the advancing forward front of the swarm midway along the breadth of the front on June 26 (Fig.~\ref{Fig:Circumstances}c), diagonally cutting a path across the swarm, and passing over the trailing and inner "corner" of the TSC near July 14 (Fig.~\ref{Fig:Circumstances}d).  By August 1 the swarm is outside the Earth's orbit into the night sky with rapidly decreasing apparent magnitude (Fig.~\ref{Fig:Circumstances}e), and by 22 August, any H=24 objects have faded from visibility at $m>26$ (Fig.~\ref{Fig:Circumstances}f). A link to an animation illustrating the TSC encounter can be found in the Supplementary Material at \url{http://www.astro.uwo.ca/~dclark56/tauridswarm/arxiv/}.  

\section{Viewing Opportunities and Constraints}
\begin{table}
	\centering
    \caption{A calendar of dates referred to in the text relevant to observation opportunities of the TSC in 2019. JD is Julian date less $2458000.5$}
    \label{Tab:Calendar}
    \setlength{\tabcolsep}{4pt}
    \begin{tabular}{cp{6.5cm}}
        \hline
        Date(JD) & Description \\
        \hline
        06/17(651) & Full moon \\
        06/17(651) & First objects are to exceed solar elongation of $60\degree$ at $m=25$ \\
        06/22(656) & 5 days after full moon \\
        06/24(658) & TSC begins passing beneath the Earth \\
        06/24(658) & Fast moving $m=22.5$ may be appear in images \\  
        07/05(669) & Beginning of viewing opportunity A with $m=22$ objects visible with sky motion less than $2.25\degree/day$ \\
        07/11(675) & 5 days prior to full moon, end of viewing opportunity A \\
        07/16(680) & Full moon (and lunar eclipse) \\
        07/21(685) & 5 days after full moon, beginning of viewing opportunity B with dimmer objects visible with low sky-motion \\
        08/10(705) & 5 days prior to full moon, end of viewing opportunity B \\
        08/15(710) & Full moon \\
        \hline
    \end{tabular}
\end{table}
\begin{figure}
    \centering
    \fbox{\includegraphics[width=8cm]{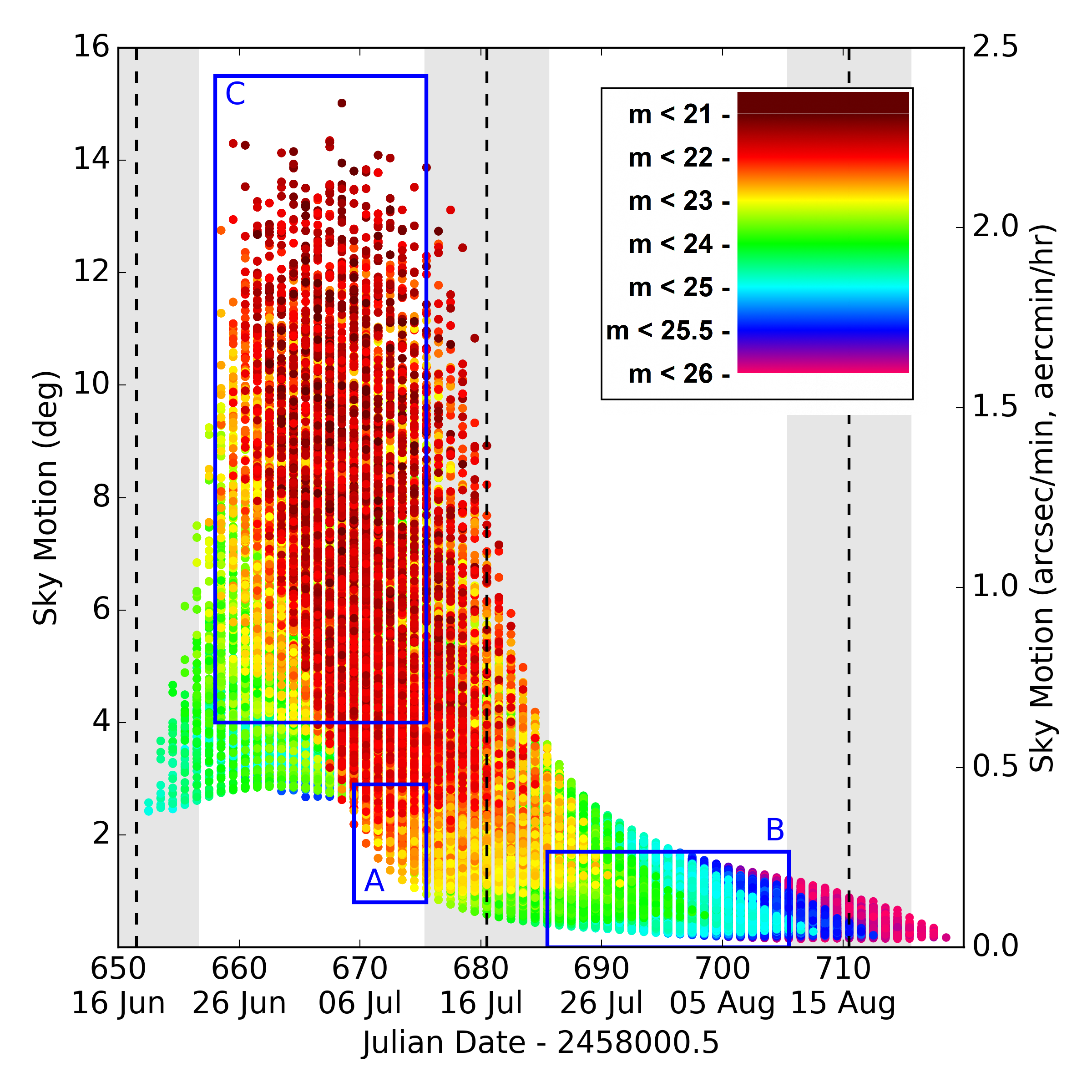}}
    \caption{TSC in-sky motion in $\degree$/day and arcmin/hour (or arcsec/min) by Julian date. All TSC objects have an assumed $H=24$.  Object colours represent apparent magnitudes corresponding to the legend. Objects with solar elongation $<~60\degree$ and objects of $m>26$ are not shown. Dashed lines represent the Full Moons with the grey representing 5 days on either side representing an approximate $60\degree$ lunar elongation potential for observation interference.  The boxed areas highlight areas of: A) relatively slow moving objects of $m<23$, B) the slowest moving objects at $m<25$ and $m<25.5$, and C) extremely bright fast moving objects.} 
    \label{Fig:MagAndMotion}
\end{figure}
Although the TSC will pass in close proximity to the Earth, observation planning must navigate the combined challenges of high sky motion, lunar phase cycles, and interference of the galactic plane.  Fig.~\ref{Fig:MagAndMotion} plots all TS object sky motion an magnitudes by date throughout the TSC passage, with three viewing opportunities (labelled A, B, C) suggested by inspection of in-sky motion, apparent magnitudes, and occurrences of a full moon on 2019/06/17 (JD~2458651.5), 2019/07/16 (JD~2458680.5), and  2019/08/15 (JD~2458710.5). The opportunity regions correspond to where objects are brightest (region C),  have the lowest on-sky motions (region B), and where both these characteristics are balanced (region A). Periods of 5 days on either side of a full moon are shaded as less favourable. Refer to Table~\ref{Tab:Calendar} for a summary of all relevant dates.  

\begin{figure}
    \centering
    \fbox{\includegraphics[width=8cm]{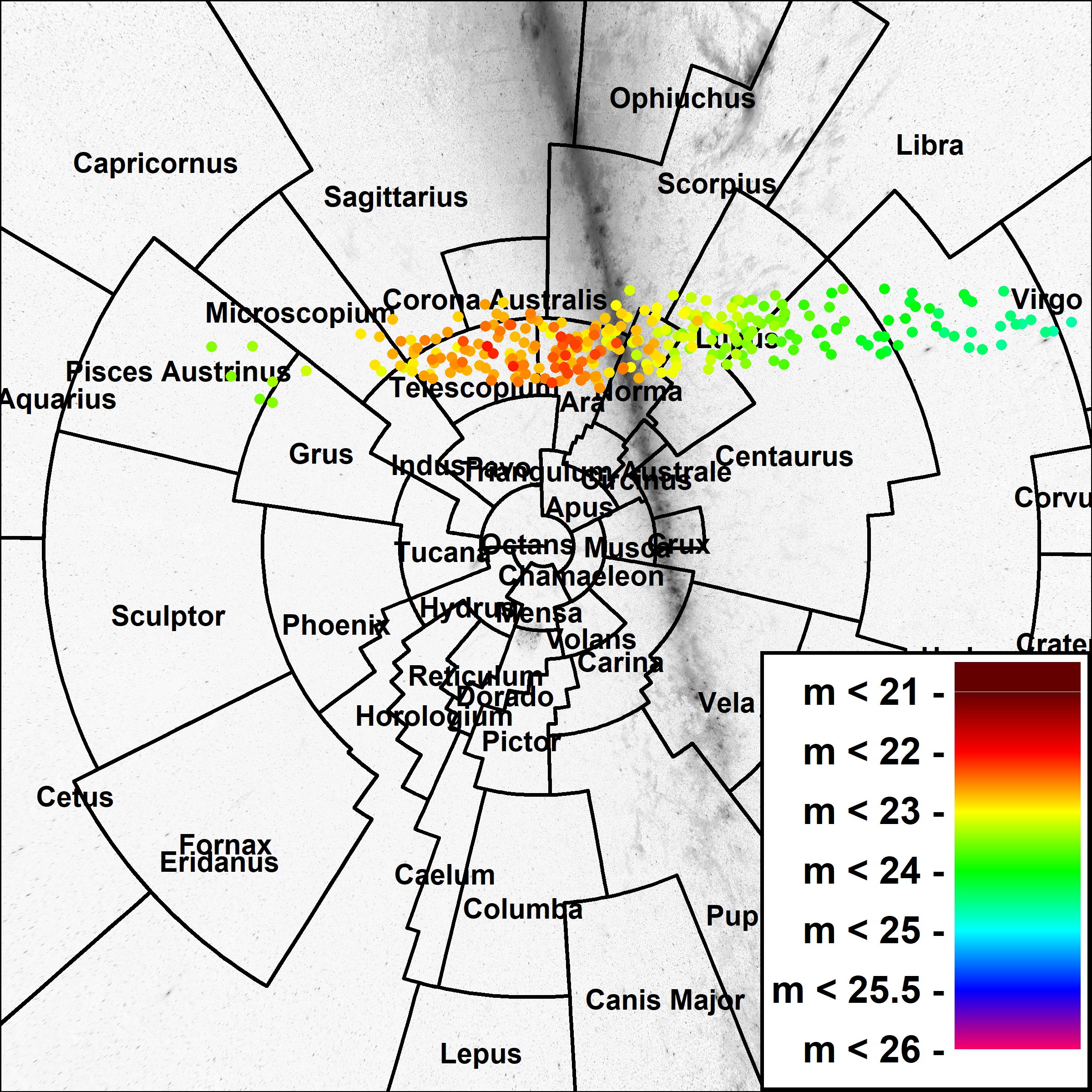}}     
    \caption{Viewing Opportunity A objects on July~11 (JD~2458675.5) with sky motion $<2.25\degree/day$ cutting a wide swath across the southern sky.}
    \label{Fig:OpportunityA}
\end{figure}
\begin{figure}
    \centering
    \fbox{\includegraphics[width=8cm]{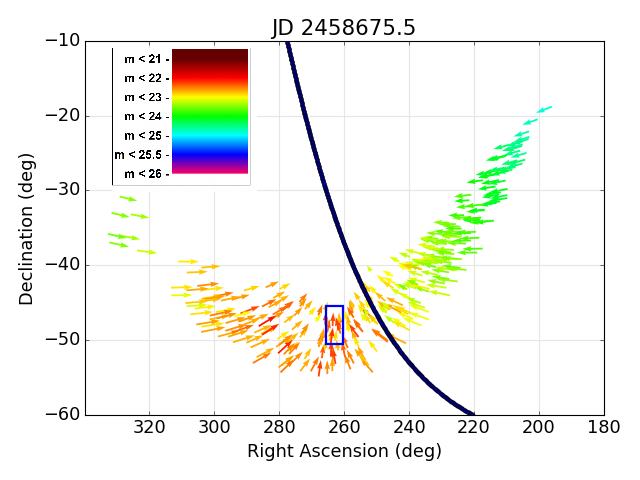}} 
    \caption{The Right Ascension and Declination of objects with sky motions of less than $2.25\degree/day$ and their direction of sky motion on July~11 (JD~2458675.5). Arrows represent one day's motion. All TSC objects have an assumed $H~~24$. Object colours represent apparent magnitudes corresponding to the legends.  The blue curved line represents the galactic plane. The blue rectangle denotes the suggested optimum observing location.}
    \label{Fig:OpportunityAMotion}
\end{figure}
Opportunity A corresponds to the date range 2019/07/05 - 2019/07/11 (JD~2458669.5 - 2458675.5) selected as a period prior to full moon where the Earth's passage above the TSC provides bright $m~~22$ objects at relatively slow sky motions as compared to the remainder of the TSC.  The upper sky motion limit of 2.25$\degree/day$ is used to capture the full date range in which $m~~22$ objects (again assuming H=24, smaller H's will of course be brighter) are visible.  Although resulting in rapid image trailing, the limit of 2.25$\degree/day$ is well within NEO survey detection limits.  The TSC location at the midpoint of Opportunity A is depicted in Fig.~\ref{Fig:Circumstances}d).  Fig.\ref{Fig:OpportunityA} illustrates this collection of objects spanning $130\degree$ of right ascension and $35\degree$ of southerly declination.  The optimum target location for detecting the brightest slow movers is in the area of $\alpha=263\degree$, $\delta=-48\degree$ (See Fig.~\ref{Fig:OpportunityAMotion}). This target area is close to the galactic plane where higher star densities hamper attempts to detect moving objects. 

\begin{figure}
    \centering
    \fbox{\includegraphics[width=8cm]{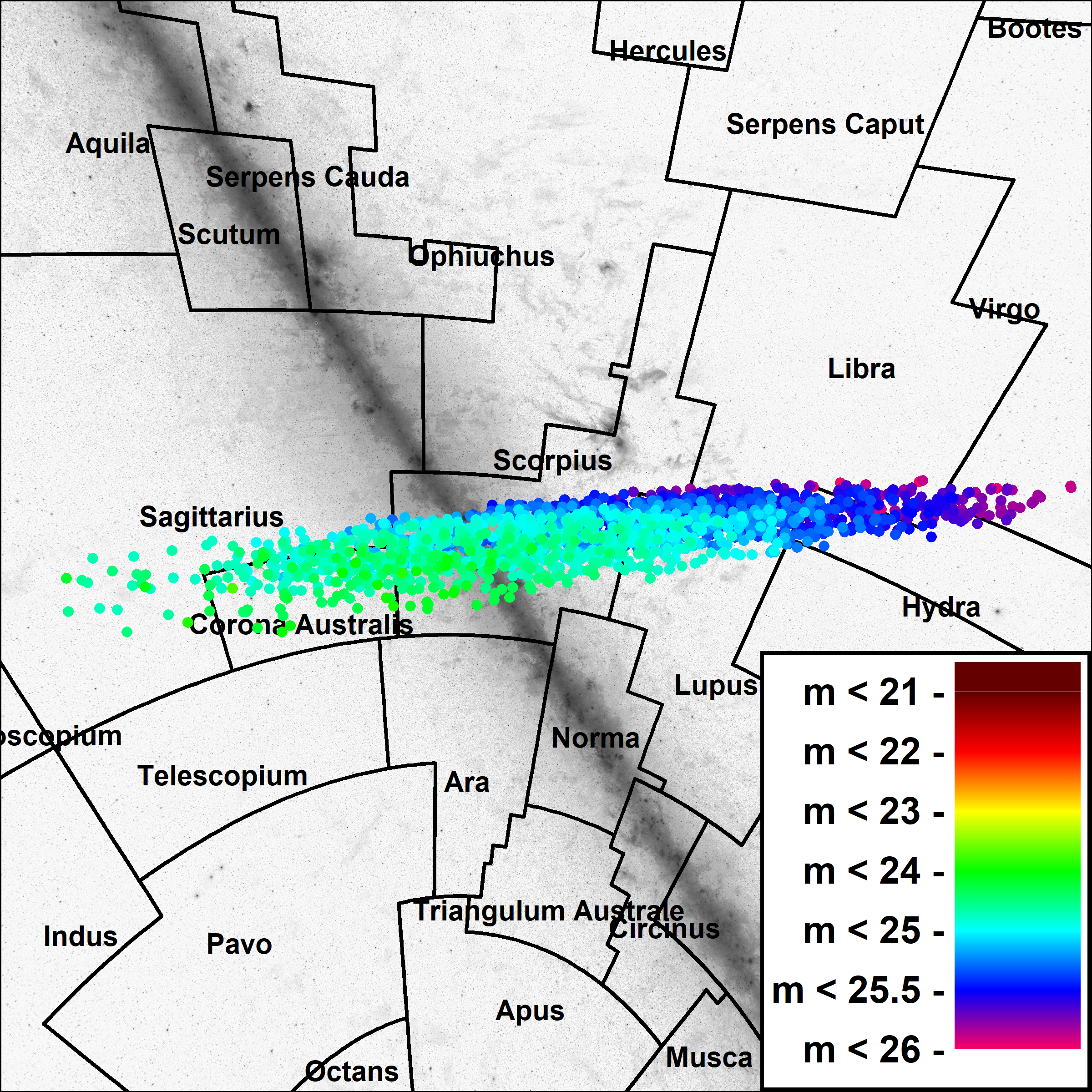}} 
    \caption{Visible TSC objects with $m<26$ at 2019/08/01 (JD~2458696.5) corresponding viewing opportunity B in Fig.~\ref{Fig:MagAndMotion}.  Objects are converging in front of the centre of the Milky Way. All TSC objects have an assumed $H~~24$. Object colours represent apparent magnitudes corresponding to the legend.}
    \label{Fig:OpportunityB}
\end{figure}
\begin{figure}
    \centering
    \fbox{\includegraphics[width=8cm]{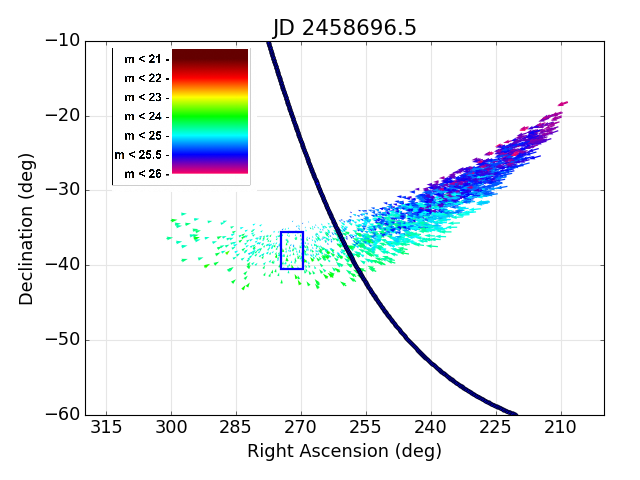}} 
    \caption{The Right Ascension and Declination of objects with in sky motion of less than $2.25\degree/day$ and there direction of sky motion on August~01 (JD~2458696.5). Arrows represent one day's motion. All TSC objects have an assumed $H~~24$. Object colours represent apparent magnitudes corresponding to the legends. The blue curved line represents the galactic plane. The blue rectangle denotes the suggested optimum observing location.}
    \label{Fig:OpportunityBMotion}
\end{figure}
Opportunity B corresponds to the date range 2019/07/21 - 2019/08/10 (JD~2458685.5 - 2458705.5) being the full observing period between the July and August full moons (See Fig.~\ref{Fig:MagAndMotion}). The TSC location at the midpoint of Opportunity B is depicted in Fig.~\ref{Fig:Circumstances}g). During this opportunity the TSC is receding from the Earth due to both the swarm's and Earth's orbital motion.  Apparent magnitudes will have increased (July 21: $m>22.5$, Aug 10: $m>25$ but are still well within the realm of larger instruments. On-sky motions will have decreased substantially to a range of near $0$ to $2\degree/day$. As with the earlier observing opportunity, the swarm continues to converge towards the galactic plane (see Fig.~\ref{Fig:OpportunityB}), unfortunately also now near the galactic centre.  Staying well east of the galactic centre and selecting the brighter objects with less sky motion (see Fig.~\ref{Fig:OpportunityBMotion}), the area around $\alpha=272\degree$, $\delta=-38\degree$ appears optimal.

\begin{figure*}
    \centering
    \fbox{\includegraphics[width=5.6cm]{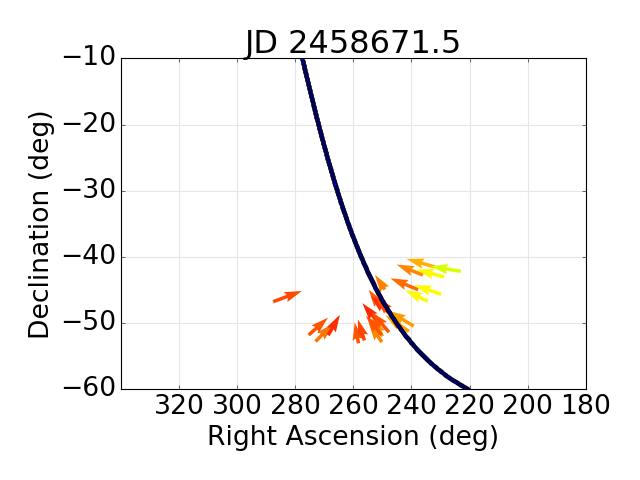}}
    \fbox{\includegraphics[width=5.6cm]{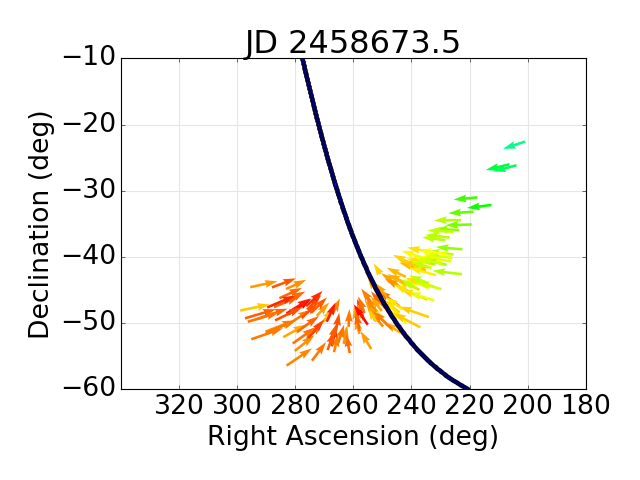}}
    \fbox{\includegraphics[width=5.6cm]{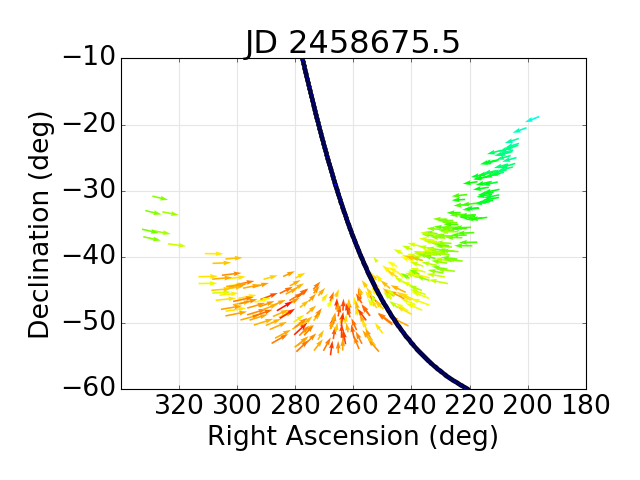}}
    \fbox{\includegraphics[width=5.6cm]{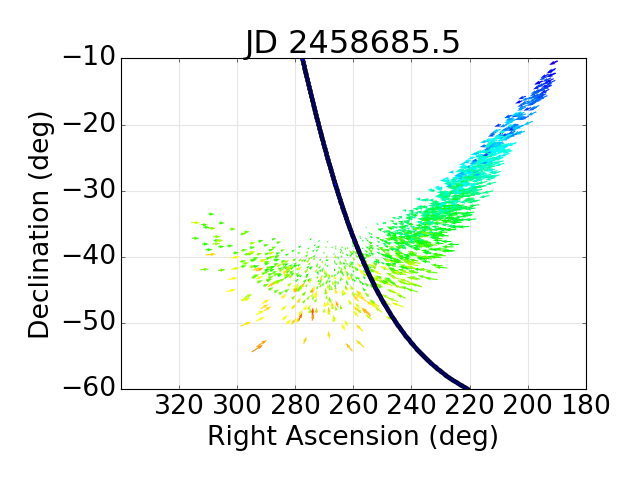}}
    \fbox{\includegraphics[width=5.6cm]{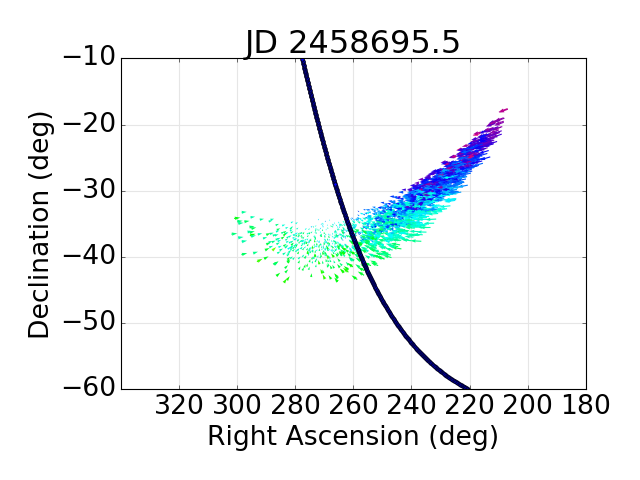}}
    \fbox{\includegraphics[width=5.6cm]{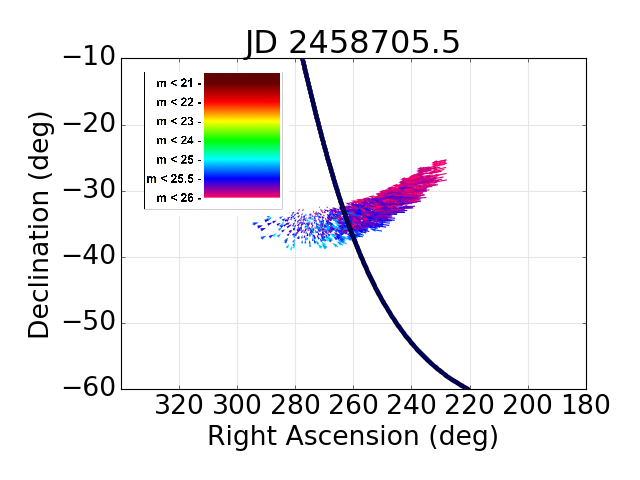}}
    \caption{The on-sky progression of the TSC through observing opportunities A (first 3 panes, in 2-day increments) and B (last 3 panes, in 10-day increments) with with arrows representing one day of on-sky motion.  Higher resolution plots including plots for intervening epochs are found in the Supplementary Material.}
    \label{Fig:MotionMontage}
\end{figure*}

Fig~\ref{Fig:MotionMontage} illustrates the progression of the TSC through the sky in July and August.  Of note is the steady movement northward and lower on-sky motion as the objects dim. Animations and higher resolution plots of the TSC position over time can be found in the Supplementary Material at \url{http://www.astro.uwo.ca/~dclark56/tauridswarm/arxiv/}.

\begin{figure}
    \centering
    \fbox{\includegraphics[width=8cm]{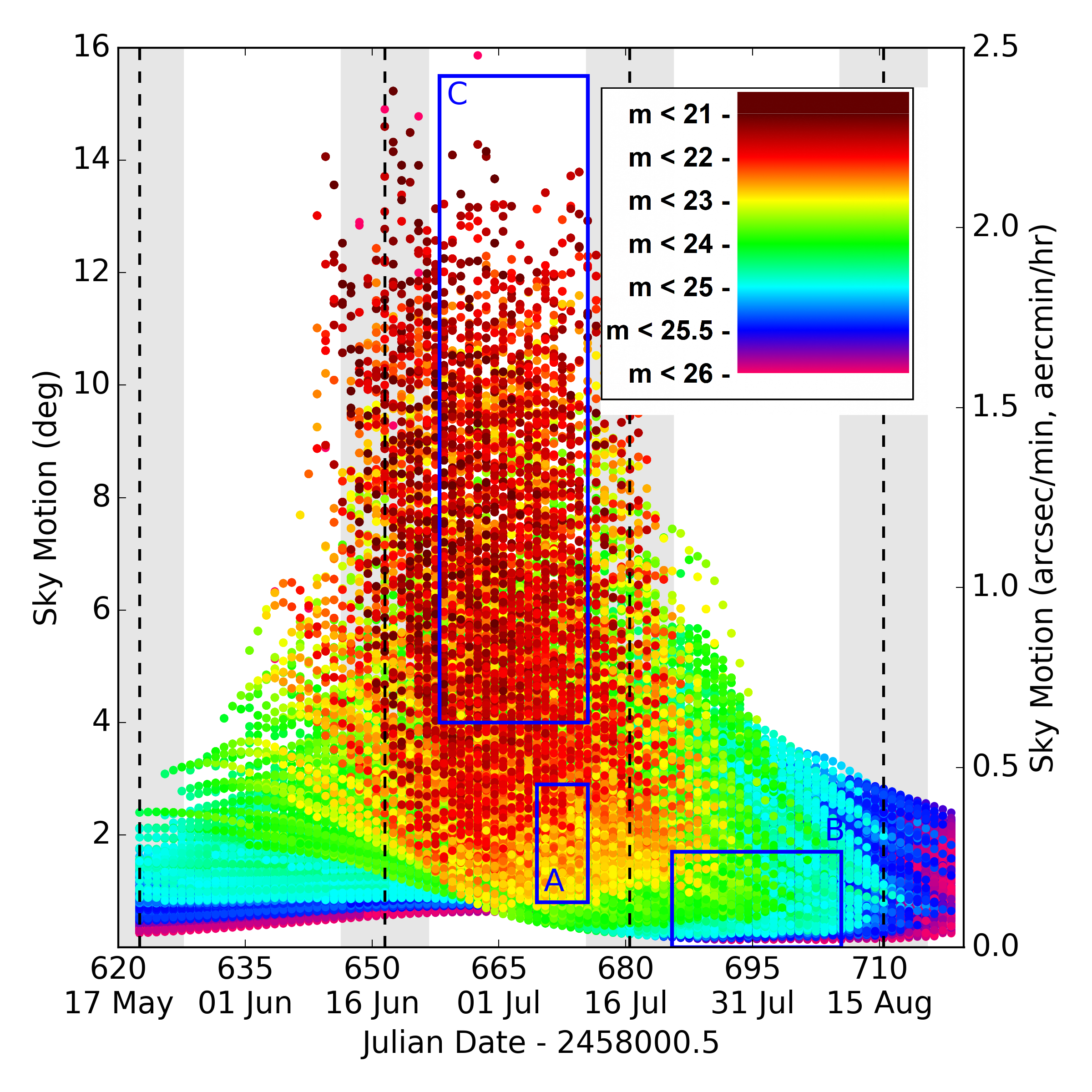}}
    \caption{The full TS in-sky motion in $\degree$/day and {arcmin/hour (or arcsec/min) by Julian date. All TS objects have an assumed $H=24$.  Object colours represent apparent magnitudes corresponding to the legend. Objects with solar elongation $<~60\degree$ and objects of $m>26$ are not shown. Dashed lines represent the Full Moons with the grey representing 5 days on either side representing an approximate $60\degree$ lunar elongation potential for observation interference.  The boxed areas highlight areas of observation opportunity shown in Fig.~\ref{Fig:MagAndMotion} and described in the text.}} 
    \label{Fig:MagAndMotion60}
\end{figure}

As mentioned in section 2, we have concentrated on a core of the TS whose members exhibit a resonance amplitude of less than $6\degree$. The full hypothesised TS includes members with resonance amplitudes out to $60-80\degree$. Fig.~\ref{Fig:MagAndMotion60} illustrates the in-sky motion of swarm members with up to $60\degree$ resonant amplitude overlaid with the viewing opportunities shown in Fig.~\ref{Fig:MagAndMotion} and described in the text.  The significant elongation of the full cloud presents additional viewing opportunities, prior to the full moon of June and after the full moon of August.  When considering the expected lower cloud density and the decreased apparent magnitude at similar in-sky motions, neither appear better than the opportunities we have identified for the TSC. 

\section{Conclusions}
The June-August 2019 encounter of the TSC provides a unique opportunity to identify additional NEOs of the swarm, helping to either substantiate or refute the giant comet hypothesis of \cite{Clube1984} and the Taurid Complex hypothesis of \cite{Asher1993a} and \cite{Asher1993a}. Dedicated surveys will at the very least be able to place limits on the NEO density near the swarm centre. The encounter will yield two opportunity windows of observation: A) 2019/07/05 -2019/07/11 (JD 2458669.5 - 2458675.5) where southern observatories may detect $m~~22$ (all apparent magnitudes assume H=24) objects with in-sky motion less than $2.25\degree/day$, and B)2019/07/21 - 2019/08/10 (JD 2458685.5 - 2458705.5) where both northern and southern observatories may view dimmer objects ($m>22.5$ in late July, $m>25$ in early August) with in-sky motion ranging from near $0$ to $2\degree/day$.  The recommended target sky position for TS object detection is in the southern sky just east of the galactic plane, $\alpha=263\degree$, $\delta=-48\degree$ for opportunity A, $\alpha=272\degree$, $\delta=-38\degree$ for opportunity B.
\section*{Acknowledgements}
We thank Dr. Auriane Egal for her contribution to the understanding and assistance in modelling of the TS. We thankfully acknowledge Mark Boslough's contributions to the recognising of possible observation opportunities to better quantify air-burst risk and to test the Taurid Complex hypothesis. We thank David Asher for his review of this article and his many helpful suggestions.

Funding for this work was provided through NASA co-operative agreement NNX15AC94A, the Natural Sciences and Engineering Research Council of Canada (Grants no. RGPIN-2016-04433 \& RGPIN-2018-05659) and the Canada Research Chairs Program.




\bibliographystyle{mnras}
\bibliography{mybibfile} 




\bsp	
\label{lastpage}
\end{document}